\documentclass[twocolumn]{webofc}

\usepackage[varg]{txfonts}   
\begin{document}
\title{New developments in fission studies within 
the time-dependent density functional theory framework}
%
%

\author{\firstname{Aurel} \lastname{Bulgac}\inst{1}
}

\institute{Department of Physics, University of Washington, Seattle, WA 981965-1560, USA}

\abstract{%
 We have extended significantly the microscopic description of the fission process by examining a larger set of observables. 
We extract neutron and proton numbers of fission fragments, their spins and fission fragment relative orbital angular 
momentum and their correlations, investigate neutrons emitted at or shortly after scission, excitation energy sharing 
mechanism, total kinetic energy of fission fragments, and the entanglement entropy. 
I will present a short overview of our simulations obtained 
with two independent nuclear energy density functionals.}
\maketitle
\section{Introduction}

Nuclear fission, which was discovered in 1939 by Hahn and Strassmann~\cite{Hahn:1939} and its leading 
microscopic mechanism was described almost immediately by Meitner and Frisch~\cite{Meitner:1939}.  In spite 
of more than 80 years since a fully microscopic description of fission dynamics still eludes us. 
Most theoretical developments have been phenomenological, many of them making 
contradictory assumptions, typically not supported by a microscopic confirmation, but in the end 
reaching agreement with data, largely due to a large number of phenomenological parameters. 
There exist as well a range of microscopic approaches, based on unchecked 
assumptions and numerical approximation. There are only three possible avenues to follow 
for a fully microscopic treatment of this non-equilibrium quantum process: 
1) solve the time-dependent many-body Schr{\" o}dinger equation (TDSE);
2) solve the time-dependent density functional theory (TDDFT) extended to nuclear systems; 
3) solve a time-dependent formulation of QCD. 
Solving TDSE is unfeasible, and moreover we do not know with enough accuracy 
the interactions between nucleons. In this respect the QCD route is even more hopeless, 
as it is even more complicated than solving TDSE. The only route to follow for perhaps many 
decades is the TDDFT and its needed extension, to which I will allude a little bit.

\section{Why use Density Functional Theory?}

It has been mathematically proven than TDDFT~\cite{HK:1964,Kohn:1965fk,Runge:1984,Dreizler:1990lr,Gross:2006,Gross:2012} 
is equivalent to the TDSE at the level of one-body 
density.~\footnote{Nuclei, similarly to electrons, need in principle an external one-body potential for this statement 
to be in agreement with Hohenberg and Kohn theorem~\cite{HK:1964}. Unlike infinite systems, in case of finite nuclei one can always add 
a one-body square-well potential of sufficiently large diameter,
which can eventually be taken to infinity, and establish thus formally the one-to-one relation 
between the number density and the many-body wave function. } 
TDDFT has the great advantage, that unlike many phenomenological approaches, is a 
quantum framework, but also a  disadvantage, as we do not know with great accuracy its main ingredient, 
the nuclear energy density functional (NEDF). The  quality of  {\it ab initio} nucleon interaction 
approaches~\cite{Marino:2021,Hu:2022} and the corresponding derived NEDFs are 
still of insufficient accuracy for the treatment of heavy nuclei. For the time being we will have to rely 
on accurate phenomenological NEDFs~\cite{Shi:2018}. 

\section{Crucial theoretical input} 

The crucial question in addressing a microscopic description of fission is: 
What are the essential ingredients without which we are doomed to fail? Surprisingly, the number of essential 
ingredients to describe accurately nuclear masses, charge radii and many other static properties of nuclei~\cite{Shi:2018}  
are exactly the same which are required for the dynamic description of fission, 
which is a time-dependent process and intrinsicaly a non-equilibrium one as well.
\begin{itemize}

\item Nuclear surface tension and Coulomb interaction between protons. These two nuclear properties are 
known from Bethe-Weizs{\" a}cker mass formula, suggested by Gamow in 1930. 
Meitner and Frisch~\cite{Meitner:1939} realized that fission is driven by the competition between the surface and 
the Coulomb energies, and is not a tunneling process as was the prevailing attitude at the time. 

\item The strength of the spin-orbit interaction. In the absence of spin-orbit interaction the average masses of 
the heavy and light fission fragments (FFs) and the presence of the second fission isomers cannot be 
described~\cite{Brack:1972,Bjornholm:1980}, and the most probable split with be two FFs of equal masses.

\item The strength of the neutron and proton pairing. In the absence of pairing fission would be strongly 
hindered~\cite{Bertsch:1980,bertsch:1997,Bertsch:2017,Bulgac:2019b,Bulgac:2020}. Inclusion of pairing requires 
an extension of DFT to superfluid systems, 
the time-dependent superfluid local density approximation (TDSLDA)~\cite{Bulgac:2013a,Bulgac:2019} in the spirit of the 
Kohn-Sham LDA~\cite{Kohn:1965fk}.

\item Symmetry energy of symmetric nuclear matter, needed to correctly describe the N/Z composition of FFs,
and to a lesser extent its density dependence.

\item Saturation density and binding energy of symmetric nuclear matter, needed to correctly describe the 
dynamics of an incompressible fissioning liquid drop~\cite{Meitner:1939,Bohr:1939}. 

\end{itemize} 

As in the case of the many-body Schr{\" o}diner equation, where we still do not know with sufficient accuracy 
the interaction between nucleons,  TDSLA suffers from a similar deficiency, an absence of a very accurate NEDF. 
There latest efforts to produce {\it ab initio} NEDF~\cite{Marino:2021} 
lead to NEDF with quite large errors in the binding energy of nuclei, as even the most advanced {\it ab initio} approaches
fail to accurately predict the properties of heavy nuclei yet~\cite{Hu:2022}. It will take a long time until the accuracy 
of the phenomenological NEDFs could be matched by {\it ab initio} methods. In this respect the situation in nuclear physics is similar 
to the situation of the Standard Model (SM) of elementary particles, which still depends of a very large  number of 
phenomenological parameters. On the other hand we can be certain that the DFT theoretical framework is otherwise correct. 
An accurate NEDF depends only on 7 phenomenological parameters~\cite{Shi:2018}, as compared with 19 in the case of SM.

\section{Fission within TDSLDA}

Presently, the microscopic description of fission can only cover the dynamics from the outer fission barrier until 
the FFs are well spatially separated. The formation of a compound nucleus and its evolution up to the outer fission 
barrier is a very long process, $\approx 10^{5...6}$ fm/c, too long to be numerically described microscopically 
with available computer capabilities within the next decade or so at least.

The goal of a fully microscopic approach to fission dynamics is to have as the only input: 1) a nuclear energy density functional (NEDF), 
which is fully characterized by the listed above essential nuclear ingredients; 2)  the proton and neutron numbers of the 
compound nucleus; 3) its excitation energy and initial spin. From these ingredients alone
one should be able to predict both some experimentally accessible observables as 
well as a number of FF properties, which are unlikely ever to be measured, but relevant for phenomenology or the
origin of elements in the Universe: 
total kinetic energy (TKE) and the total excitation energies (TXE) of the FFs and its sharing between FFs, 
the FF proton and neutron numbers, FF spin and parities, FF excitation energies  before the emission of 
neutrons and statistical gammas, the mass and charge yields, 
and hopefully also the correlations between various observables.  Once these FF properties 
have been extracted they can be further used in post-processing approaches such as
CGMF~\cite{Becker:2013}, FREYA~\cite{Vogt:2009} FIFRELIN~\cite{Litaize:2012} 
to further improve them and to extract neutron multiplicities and gamma spectra.

\begin{figure}
\includegraphics[width=1.1\columnwidth]{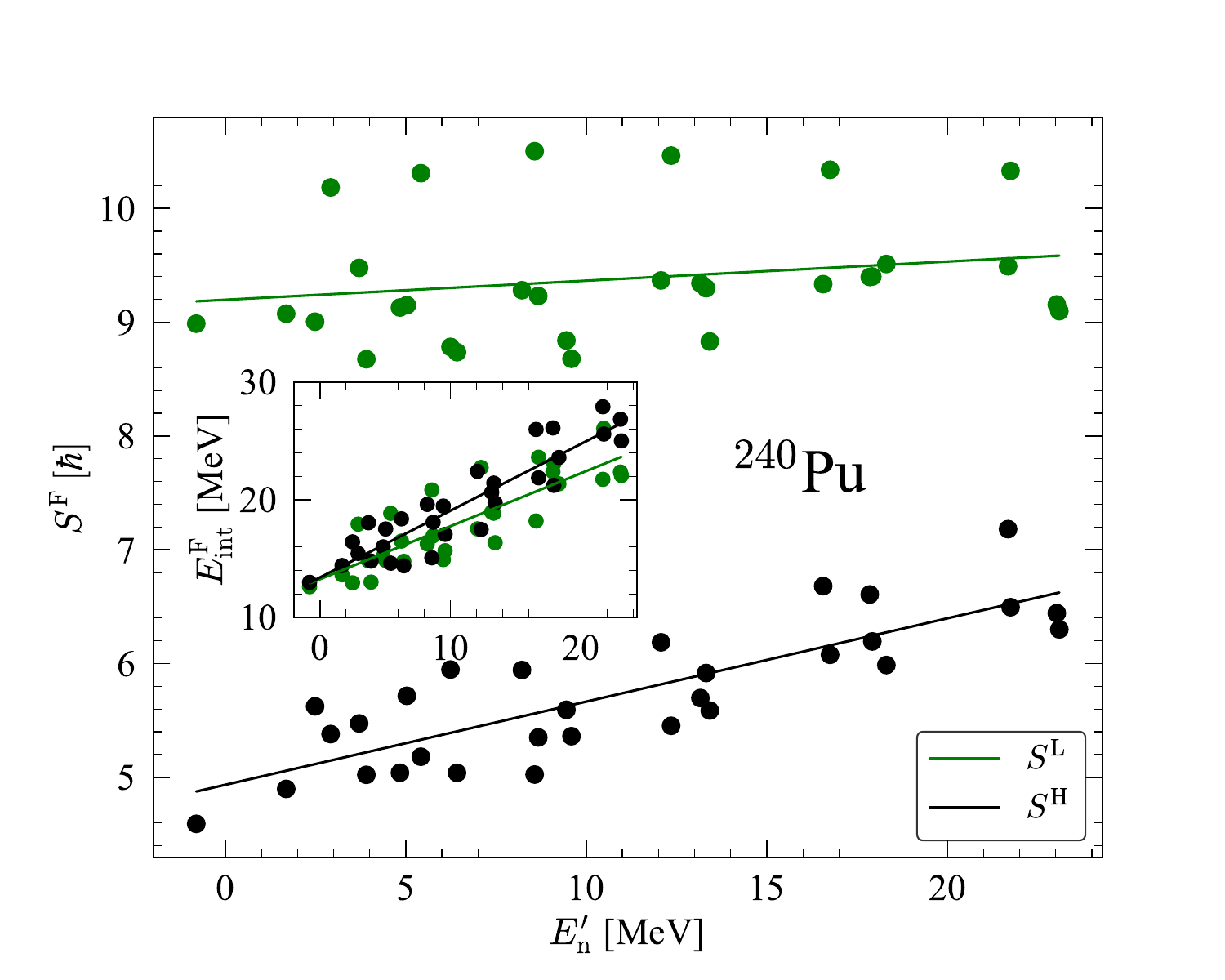}
\caption{ \label{fig:EvsJ} (Color online) 
The average intrinsic spins $S^\text{L,H}$ versus the initial FF equivalent incident neutron energy $E_n'= E^*-S_n$ 
($E^*$ and $S_n$ are the excitation energy and $S_n$ the neutron separation energy)
for the reaction 
$^{239}$Pu(n,f) with SkM$^*$ NEDF. The solid 
lines are linear fits over the data, $S^\text{L}=0.0168\,E_n'+9.197$ and $S^\text{H}=0.0732\, E_n'+4.933$ respectively,
as a function of equivalent neutron energy $E_n'$ along with their linear fits.
In the inset we display the FF excitation energies and their linear fits $E^\text{L}_\text{int}=0.4505 \, E_n'+ 13.25$ and 
$E^\text{H}_\text{int}=0.5676\,E_n'+13.40$. Using 
$E^\text{F}_\text{int}\approx A^\text{F}(T^\text{F})^2/10$~\cite{Bulgac:2019b,Bohr:1969} it follows that on average $T^\text{L}>T^\text{H}$. }
\end{figure}

The non-equilibrium character of fission dynamics is a well  established experimental fact. 
In induced fission $^{235}$U(n$_{th}$,f)~\cite{Madland:2006} 
one starts with relatively cold quantum system, where pairing correlations, 
at the top of the outer fission barrier, are essential ingredients, 
as has been established by theory since 1970s.
The average TKE of the FFs is  $\approx170-175$ MeV, 
while the mass difference between the compound nucleus and final products is
$\approx 210$ MeV.  As a result the FFs end up, before emitting 
any prompt neutrons or gammas with roughly 40 MeV of 
excitation energy to share among them. 
Therefore, the FFs 
emerge hot from an initial cold compound nucleus, 
with the heavy FF (HFF) colder than the light FF (LLF), see also Fig.~\ref{fig:EvsJ}.   
The assumption that a nucleus is a mostly incompressible liquid drop, can be easily quantified
within  the fully quantum framework TDSLDA, for which the total energy $E_{tot}$ is conserved.
As for any liquid in motion,  the total energy can be uniquely separated into two parts
\begin{eqnarray} 
E_{tot} = E_{int}(t) + E_{flow}(t),\\
E_{flow}(t) = \int d^3r \frac{{\bf j}^2({\bf r},t)}{2m n({\bf r},t)}
\end{eqnarray}
where $n({\bf r},t)$ is the nucleon number density, ${\bf j}({\bf r},t)$ is the collective 
momentum flow of the nuclear fluid, $m$ the nucleon mass, and $E_{int}(t)$ the part of the nucleus 
energy which depends only on the matter distribution. In the only available microscopic evaluation
of $E_{flow}(t)$~\cite{Bulgac:2019,Bulgac:2020}  it was demonstrated that 
a fissioning nucleus, while descending from the top of 
the outer barrier on the the way to the scission configuration, has an almost negligible 
collective flow energy of the order of 1-2 MeV, in spite of the fact that the energy difference in 
numerous calculated collective potential energy surfaces  shows a gain of $\approx 20$ MeV. 
The microscopic collective potential energy surfaces are evaluated by minimizing the total energy 
of a nucleus enforcing various in the presence of various 
quadrupole, octupole, hexadecapole, etc. deformation constrains~\cite{Marevic:2022}, 
therefore assuming that while evolving towards scission a nucleus is always at zero intrinsic temperature. 
While sliding down towards scission the nuclear shape evolves very slowly, as that of an extremely viscous 
fluid, with a collective speed almost an order of magnitude smaller that in an adiabatic shape 
evolution of the nuclear shape. The nuclear shape evolution is clearly an irreversible and a highly 
non-equilibrium process, during which the average properties of the FFs slowly emerge, though they 
are not yet  fully defined.

\begin{figure} 
\includegraphics[width=1\columnwidth]{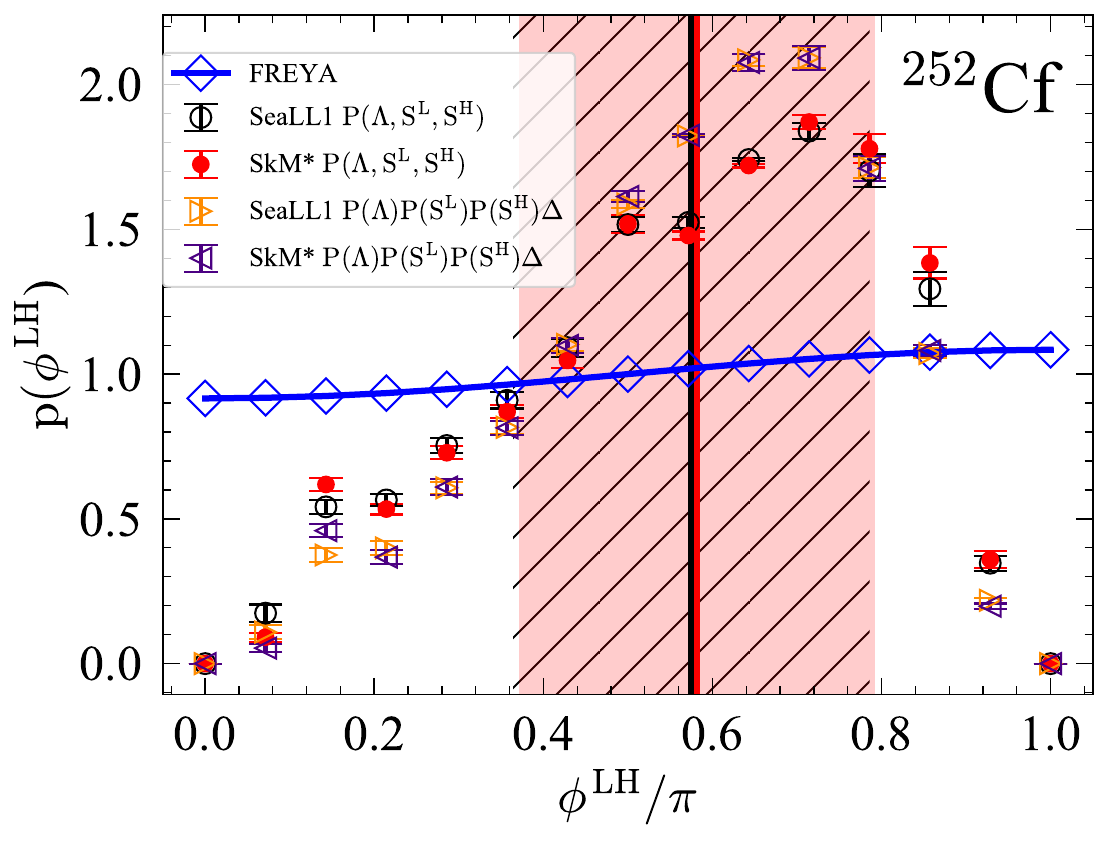}
\caption{ \label{fig:Cos} The  circles and bullets
represent the histogram (bin size = 0.22 radian) of the angle between the FF intrinsic spins ${\bf S}^\text{L}$ and ${\bf S}^\text{H}$, 
extracted using the triple distribution $P(\Lambda,S^\text{L},S^\text{H})$ obtained in Ref.~\cite{Bulgac:2022b} and illustrated in Fig.~\ref{fig:Le} 
to evaluate $p(\phi^\text{LH})$, 
$\int_0^\pi d\phi^\text{LH}p(\phi^\text{LH})=1$. 
The triangles represent the histogram obtained with $P(\Lambda)P(S^\text{L})P(S^\text{H})\Delta$, see Ref.~\cite{Bulgac:2022b}. 
 The blue line and diamonds are the prediction of the FREYA model~\cite{Randrup:2021}. 
The distributions   $p(\phi^\text{LH})$ for $^{236}$U$^*$ and $^{240}$Pu$^*$ are very similar. 
This figure is reproduced from Ref.~\cite{Bulgac:2022b}.}
\end{figure} 

Apart form microscopically firmly establishing for the first time that the large amplitude collective motion  (LACM)
in fission dynamics is strongly damped it was also shown that the TKE, TXE and average 
FF proton and neutron numbers are predicted with a roughly 1\% accuracy~\cite{Bulgac:2016,Bulgac:2019b,Bulgac:2020}, 
without the resorting to any approximations or any unchecked theoretical assumptions, such as the introduction 
of collective potential energy surface and collective inertia, which assume that LACM is adiabatic, as was 
the prevailing attitude of theorists over several decades until now~\cite{Ring:2004,Schunck:2016,Pomorski:2012}. 
The strongly damped character of LACM provides the theoretical justification for the of 
the brownian motion model~\cite{Ward:2017,Albertsson:2020,Albertsson:2021,Albertsson:2021a}. Phenomenologically, 
in order to describe the fission yields alone a plethora of models has been suggested in literature, all of them succeeding in explaining 
data, due to a sufficient number of phenomenological parameters and based of contradictory theoretically unchecked 
assumptions~ \cite{Wilkins:1976,Randrup:2011,Aritomo:2014,Sierk:2017,Ishizuka:2017,Sadhukhan:2016,
Sadhukhan:2017,Regnier:2016,Regnier:2019,Lemaitre:2015,Lemaitre:2021} (to quote a few). The only conclusion one can draw 
from these approaches is that FF yields are largely insensitive to the model and their often contradictory assumptions. 

\begin{figure}
\includegraphics[width=1\columnwidth]{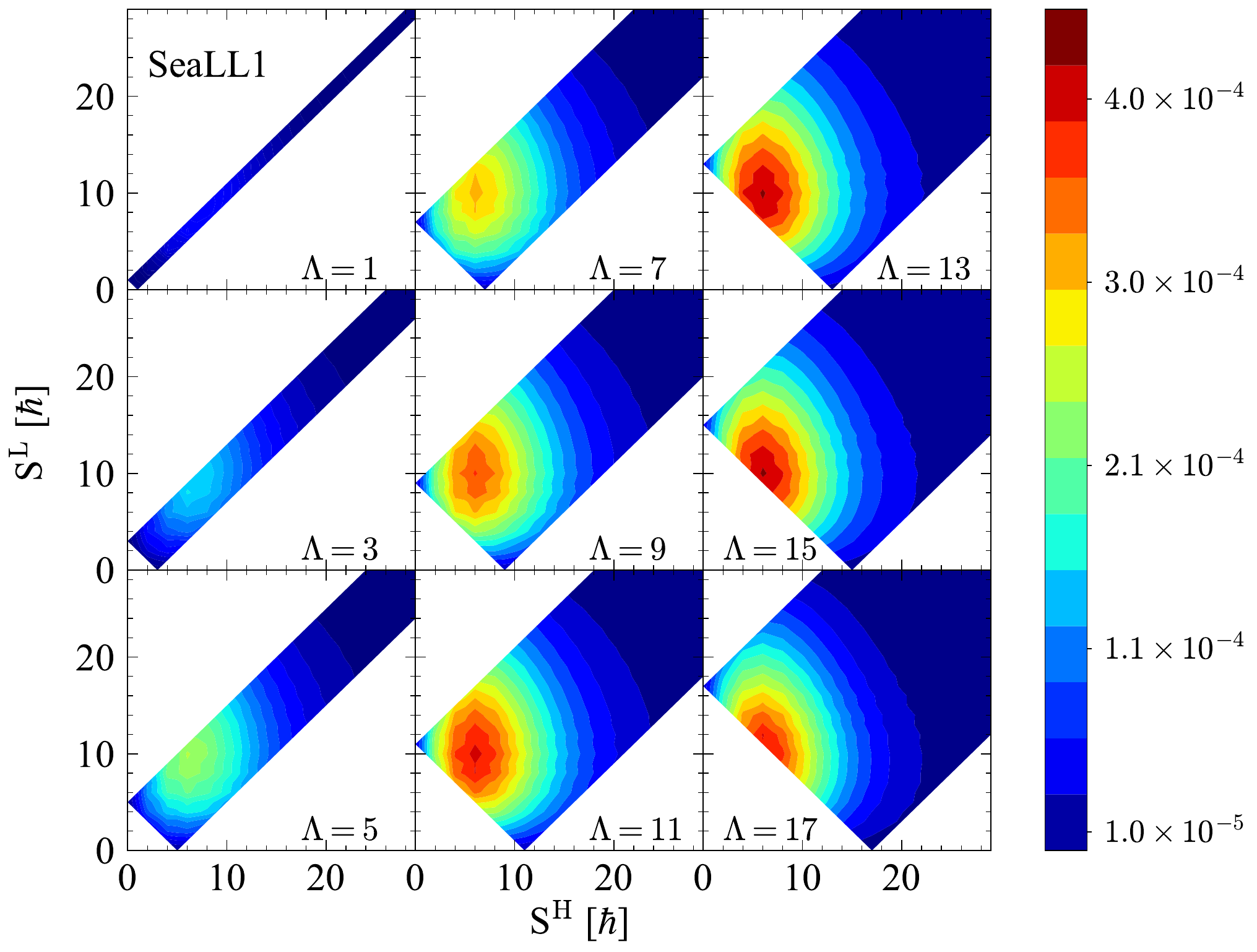}
\caption{ \label{fig:Le}The $^{252}$Cf triple probability distribution $P(\Lambda,S^\text{L},S^\text{H})$ for  SeaLL1
NEDF for odd values of $\Lambda$. 
The FF parities are correlated with the orbital angular momentum $\pi^\text{L}\pi^\text{H}=(-1)^\Lambda$. 
This triple distribution vanishes outside the region $|S^\text{L}-S^\text{H}| \le \Lambda \le S^\text{L}+S^\text{H}$, 
shown with white in these plots. The distributions for $^{236}$U$^*$ and $^{240}$Pu$^*$ are very similar. 
This figure is reproduced from Ref.~\cite{Bulgac:2022b}.}
\end{figure}  

Another aspect of the fission dynamics which proved rather difficult to interpret correctly
is to explain how the FF intrinsic spins are generated, see the recent experimental results~\cite{Wilson:2021}, 
where it was stated that the FF spins are generated after scission.  The Coulomb interaction between receding 
FFs can lead to some contribution to the final FF spins~\cite{Bertsch:2019b, Scamps:2022}, but not fully explain 
their magnitudes, distributions, and correlations. Many other phenomenological models have been 
suggested over the years and the recent workshop~\cite{workshop:2022}  can provide some 
insight into the literature. While the initial spin the compound momentum 
nucleus is very small, and zero in case of spontaneous fission of $^{252}$Cf
for example, the FFs have an average spin of the order of 10$ \hbar$~\cite{Wilhelmy:1972,Wilson:2021}
with the spin of the HFF on average smaller than the spin of 
the LFF as established by microscopic calculations~\cite{Bulgac:2021,Marevic:2021,Bulgac:2022b}. 
Fig.~\ref{fig:EvsJ} is an example of the dependence of the FF intrinsic spins 
and in the inset the excitation energy of the two FFs 
 as a function of the initial excitation energy of the compound nucleus obtained from TDSLDA simulations.

While the theoretical literature addressing the FF spins is quite extensive, see references in 
Refs.\cite{Bulgac:2021,Bulgac:2022b,Marevic:2021,Randrup:2021}, 
I will discuss here only the latest phenomenological 
approach FREYA~\cite{Vogt:2021,Marevic:2021,Randrup:2021,Randrup:2022,Randrup:2022a} used to compare 
its predictions with recent data~\cite{Wilson:2021}. In FREYA the FF spins and their relative orbital angular 
momentum are treated classically, assuming that the rotational energy 
\begin{equation}
E_\text{rot}=  
                 \frac{ {\bf S}^{L}\cdot {\bf S}^{L}} {  2I^{L} } + 
                  \frac{ {\bf S}^{H}\cdot {\bf S}^{H} }{ 2I^{H} }
                  +\frac{{\bf \Lambda}\cdot {\bf \Lambda} }{2I^\text{R}}.
                  \label{eq:rot}
\end{equation} 
It is implied in this approach that these rotational degrees of freedom are thermalized before scission, as being 
engulfed by the rest of the nucleon degrees of freedom, which form a thermal bath. In the case of $^{252}$Cf
the quantum mechanical average 
\begin{equation} 
 \langle {\bf S}^{L} + {\bf S}^{H} +{\bf 
\Lambda}\rangle  ={\bf 0} \label{eq:S0} 
\end{equation}  
should vanish at all times.
The two emerging, but  not fully defined, FFs are in contact with the bath only during the time the nucleus descend from the top of 
the outer barrier until scission. After separation each FF is isolated and its spin is conserved and before the
compound nucleus reaches the top of the outer barrier one can safely assume that no trace of an emerging FF exists.
One can easily estimate the period of rotation of an emerging FF from relation between the spin, moment of 
inertia and its angular velocity $\hbar S^F= I^F \omega^F$ and obtain that $2\pi/\omega^F \approx 3,400$ fm/c, 
assuming that $S^F\approx 10$ and $I^F$ is the moment of inertia of a rigid sphere.   Therefore the period of rotation 
of a FF while still in contact with the other FF and the thermal bath is significantly longer than the time needed for 
the compound nucleus to descend from the top of the outer barrier to scission, which is $\approx 1,500$ fm/c. 
From numerous kinetic studies of gases it is known that starting with arbitrary initial conditions one reaches
an approximate Boltzmann distribution after 3-5 collisions. A classical rigid rotor would need to arrive in 
equilibrium with the bath at an average angular momentum of $10\hbar$ in the time it barely  
performs half a rotation, thus an uncertainty in the angle $\Delta \phi \approx {\cal O}(\pi)$, and therefore a 
wave packet with an enormous uncertainty $\Delta S\Delta \phi = {\cal O}(10)$.
Whether such a classical model with such large uncertainty is realistic to describe its equilibration with the rest of the nuclear 
system or even only of its rotational modes in such a short time, it is not obvious. 
Other potential  issues with the assumption adopted in FREYA have been discussed in 
Refs.~\cite{Bulgac:2022b,Bulgac:2022d}.
 
 So far only the phenomenological model FREYA has made a prediction concerning the angle 
 between the spins of the two emerging FFs in Ref.~\cite{Vogt:2021}. Soon after that the TDSLDA microscopic treatment 
 has also made a prediction~\cite{Bulgac:2022b}, see also Ref.~\cite{Bulgac:2022d}, 
 and the two theoretical predictions are in stark disagreement with each, see Fig.~\ref{fig:Cos}. Several experimental 
 groups plan to shed light on this issue (L. Sobotka private communication).  
 
 In Refs.~\cite{Bulgac:2021,Bulgac:2022b} we have evaluated for the first time in a fully microscopic approach the
 single and triple distribution of the FF intrinsic spins and their relative orbital angular momentum and also their correlations.  
 From the triple angular momentum distribution extracted in Refs.~\cite{Bulgac:2022b,Bulgac:2022d} one can determine the single and double 
 FF distributions
 \begin{eqnarray} 
&{P}_1(S^{F} )&= \sum_{\Lambda, f\neq  F} {P}_3(S^{F},S^{f},\Lambda),\\
&{P}_1(\Lambda) &= \sum_{S^L,S^H}{P}_3(S^{L},S^{H},\Lambda),\\
&{P}_2(S^{L},S^{H})&= \sum_\Lambda {P}_3(S^{L},S^{H},\Lambda), 
\end{eqnarray} 
and evaluate 
 \begin{eqnarray}
&&\sum_{ S^{L,H},\Lambda } | {\cal N}{P}(\Lambda){P}_1(S^{L}) {P}_1(S^{H})\triangle  \nonumber \\
&&- {P}_{3}(S^{L},S^{H},\Lambda)= 0.35,\label{eq:S3}\\
&&\sum _{S^L,S^H} |{P}_2(S^{L},S^{H}) -  {P}_1(S^{L}){P}_1(S^{H})| = 0.02,\label{eq:S2}
\end{eqnarray}
 where ${\cal N}$ is a normalization constant and 
 $\triangle$ enforces Eq.~\eqref{eq:S0}.
 Eq.~\eqref{eq:S3} reveals that the two FF spins and the relative orbital angular momentum are rather strongly correlated. 
 At the same time, Eq.~\eqref{eq:S2} tells us that if one measure  $P_2(S^{L},S^{H})$
 the two FF spins appear practically uncorrelated, as observed in the recent 
 experiment~\cite{Wilson:2021}. This is in stark contradiction with the assertion made by Wilson {\it et al.}~\cite{Wilson:2021},
 and who  interpreted this absence of correlations 
 between the two FF spins as an argument that these spins are generated after the FFs are separated. 
 The FFs emerge very deformed after scission, and their shapes relax significantly after 
 scission~\cite{Bulgac:2019b,Bulgac:2020}, however the total wave function of the entire system is strongly 
 correlated and the two emerging FFs are entangled, see also below. The presence of an ``angular momentum 
 bath $\Lambda$'' appear to be sufficient to ensure the apparent independence of the two FF spins, in spite of 
 the triple strong FF correlations with $\Lambda$. 
 
\begin{figure}
\includegraphics[width=1\columnwidth]{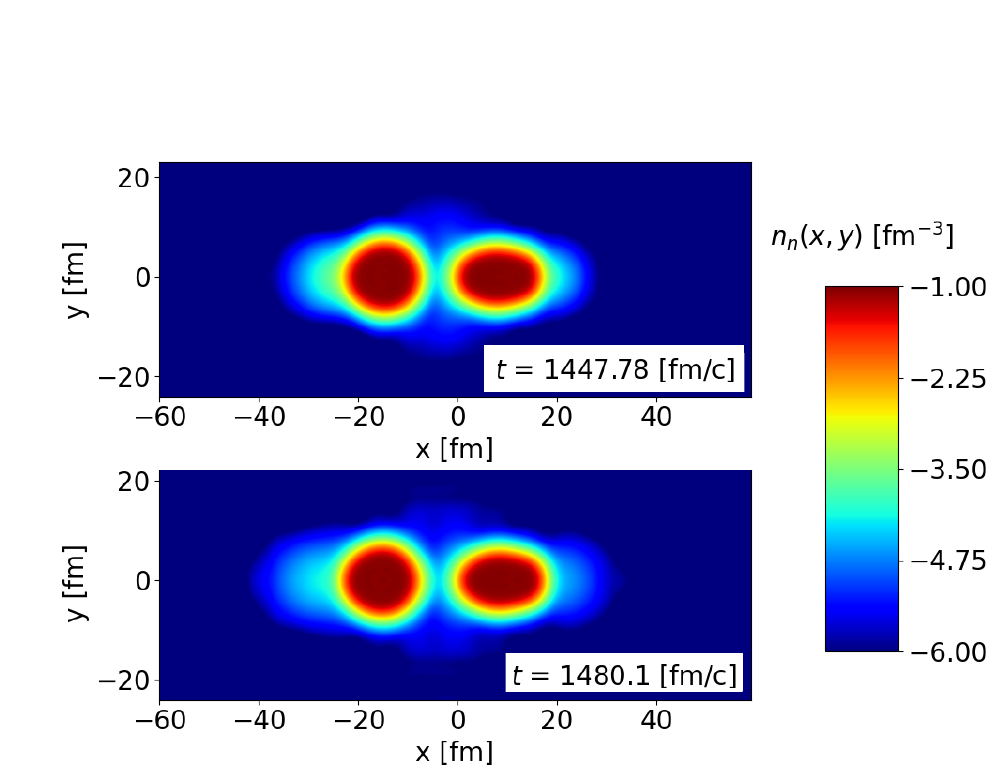}
\caption{ \label{fig:neut} Two consecutive frames of the neutron distribution 
after the two FFs separated in $^{235}$U(n,f) induced fission. 
The HFF is on the left and the LFF is on the right and the yellow band is the surface of each fragment, 
where the neutron number density is $< 0.01$ fm$^{-3}$. While one can clearly see some 
neutrons emitted from the neck, most neutrons appear to be emitted in the direction of the motion of each FF. (The colorbar shows the $\log n_n$.)}
\end{figure}    

Since we follow the two FFs until they are widely separated we can get insight into the neutron emission before 
the FFs are fully accelerated. There is a long debate in literature concerning the existence of scission neutrons 
and, more specifically neutrons emitted from the neck forming between the two emerging fragments before scission, 
see references in Ref.~\cite{Bulgac:2020a}. While the FFs are being accelerated the mean field potential experienced 
by neutrons is tilted, as in a bucket filled with water, and neutrons can escape. In Fig.~\ref{fig:neut} we show the 
neutron number density profile at two instances after scission. One can see the formation of neutron clouds 
in front of both FFs neutrons, along the fission axis. At the same time there are neutrons emitted perpendicular to the fission axis,
in the region between the two FFs, which might be considered to be emitted from the ``neck.'' As these results are still being 
analyzed and we did not accumulate enough data yet, and we are not prepared yet to provide more detailed estimates concerning
the number and spectrum of these emitted neutrons. 

\begin{figure}
\includegraphics[width=\columnwidth]{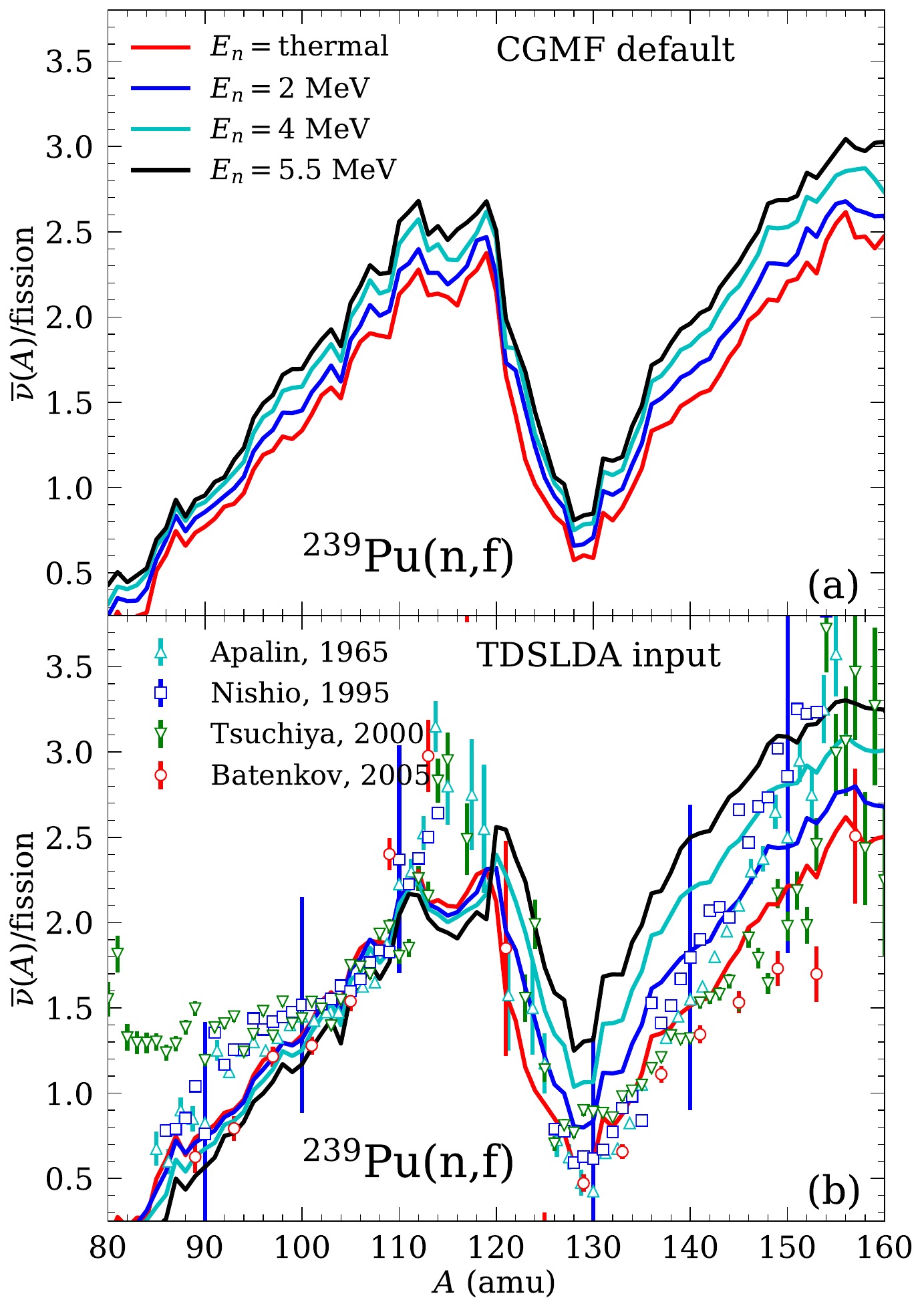}
\caption{\label{fig:nubar}
We compare in Ref.~\cite{Bulgac:2019b} the average prompt neutron multiplicity $\bar{\nu}(A)$ emitted by FFs in the case of a default 
CGMF simulation~\cite{Becker:2013}, which assumes no $E_n$ dependence for the energy sharing,  with the one extracted using the the excitation 
energy sharing between the FFs in our calculation with NEDF SeaLL1, as a function of the equivalent incident neutron energy in 
$^{239}$Pu(n,f) reaction along with available experimental data for the reaction $^{239}$Pu(n$_\text{th}$,f) from 
various experiments. The fragment mass $A$ is before neutron emission. In the GCMF default version the effect of the incident neutron energy 
is significantly milder that when TDSLDA input is used, which leads to better agreement with data.
}
\end{figure}

What happens to the FF shapes after scission, apart from the fact the the FFs recede 
very fast from one another due to Coulomb repulsion? This is an aspect which is not 
incorporated in any phenomenological FF yields studies,
even though often it is recognized, but almost never  
correctly quantified that the FFs relax their shapes. In Refs.~\cite{Bulgac:2019b,Bulgac:2020} 
it was also demonstrated that the FF shapes continue to evolve in the same manner as before the scission, 
very slowly and irreversibly, and deformation energy is converted into internal excitation energy (thermal). 
While this stage is relatively slow, it is also many order of magnitude shorter that the time scale 
an emerging FF starts emitting prompt neutrons and statistical gammas. In Fig.~\ref{fig:nubar} one can see the
neutron multiplicity spectrum obtained using the Hauser-Feshbach framework~\cite{Becker:2013} 
with input from TDSLDA is a significant improvement over what the default phenomenological GCMF predicted so far.

\begin{figure}
\includegraphics[width=0.9\columnwidth]{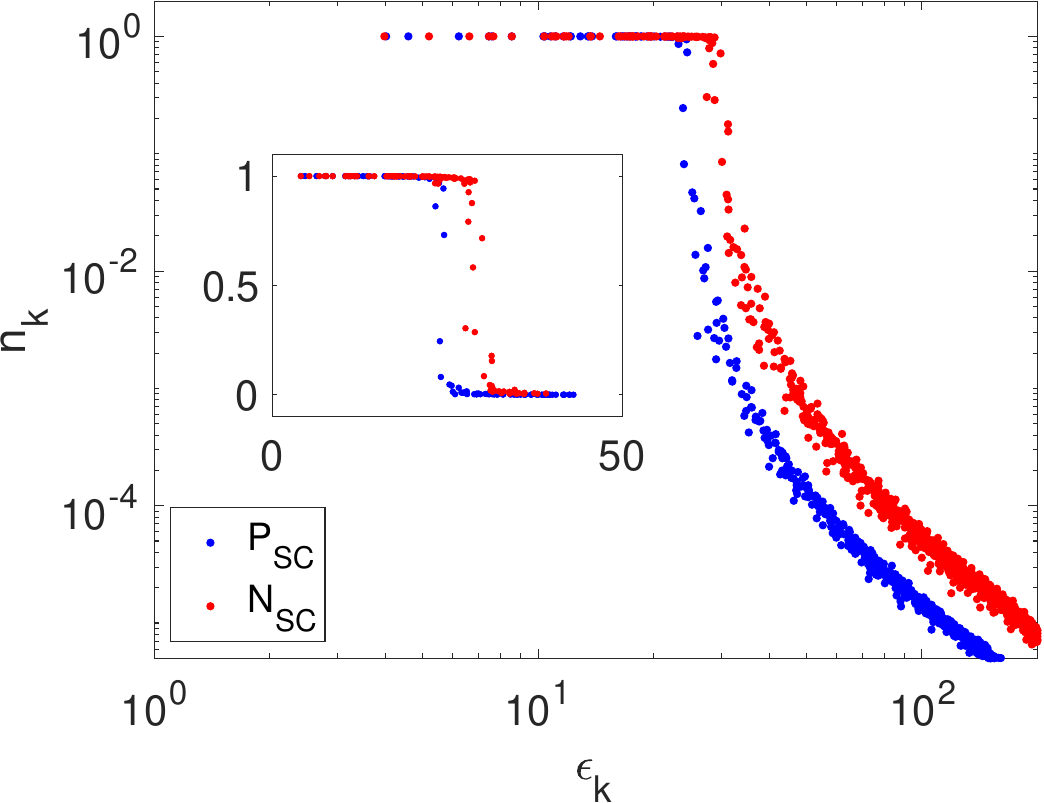}
\caption{ \label{fig:ekin}  
The canonical occupation probability $n_k$ as a function of 
$\epsilon_k=\langle \phi_k|-\hbar^2\Delta /2m|\phi_k\rangle $. 
In the inset we show that the canonical occupation probabilities $n_k$ around the Fermi level have the 
expected textbook behavior. Notice the presence of SRCs for  $\epsilon_k\propto 1/k^4 > 120$ MeV. This figure is from Ref.~\cite{Bulgac:2022c}  }
\end{figure}     

\section{High-momentum tails, entanglement entropy}
  
The TDSLDA simulations of fission are performed on a spatial cartesian lattice, with a lattice constant $l=1$ fm, 
which corresponds to a momentum cutoff~\cite{Bulgac:2013} $p_{cut}=\hbar \pi/l\approx 600$ MeV/c. This value of 
the momentum cutoff is on the upper limit considered in extracting the nucleon interactions within the $\chi$ 
Effective Field Theory, and one might expect that (some) short range correlations (SRCs) are present within TDSLDA.
Indeed, as discussed in Refs.~\cite{Bulgac:2022,Bulgac:2022a,Bulgac:2022c} $nn$ and $pp$ SRCs 
are indeed present and as expected they show up in the single-particle momentum distribution, see Fig.~\ref{fig:ekin}, 
where at momenta larger than the Fermi momentum one clearly observes that $n_k=C/k^4$, where $k$ is the wave vector and $C$ 
is Tan contact~\cite{Tan:2008a,Tan:2008b,Tan:2008c,Zwerger:2011}. Such a behavior of $n_k$ has been predicted also by 
Sartor and Mahaux~\cite{Sartor:1980} and put in evidence in experiments at the Jefferson National Laboratory~\cite{Hen:2014,Hen:2017}. 
In nuclei $np$ correlations, in particular the tensor $np$ interaction, are the dominant source of the behavior $n_k=C/k^4$, 
which are absent in TDSLDA , but which we plan to include the generalized TDSLDA~\cite{Bulgac:2022}. The $n_k$ 
long momentum tails shown in Fig.~\ref{fig:ekin} are present at all times and all excitation energies, in particular 
in the fully separated FFs, when pairing correlations are absent. In order to evaluate the momentum distribution shown 
in Fig.~\ref{fig:ekin} one has to find the eigenfunction and eigenvalues of the one-body density matrix $n(\xi,\zeta)$, 
known as canonical wave functions in superconductivity/superfluidity~\cite{Ring:2004} 
and natural orbitals, mainly in chemistry and lately also in nuclear physics~\cite{Lowdin:1955,Coleman:1963,Davidson:1972,Johnson:2018},
\begin{eqnarray}   
&n(\xi,\zeta) = \langle \Phi|\psi^\dagger(\zeta)\psi(\xi)|\Phi\rangle,  \label{eq:nntr1}\\
&\int d\zeta \, n(\xi,\zeta)\phi_k(\zeta) = n_k\phi_k(\xi),\, 0\le n_k\le 1,\label{eq:nnn1}
\end{eqnarray}
where $\xi=({\bf r},\sigma,\tau),\zeta ={\bf r}',\sigma',\tau') $     are the spatial, spin, and isospin coordinates.
The rather unexpected and surprising lesson emerging from these results is that within TDSLDA 
one can describe both long-range and short-range correlations. This a long-time
dream in theoretical physics, to generate a formalism which includes apart from the mean field evolution, as in TDHF, also the effect 
of collisions, as in the semiclassical Boltzmann-Uehling-Uhlenbeck (BUU) equation~\cite{Nordheim:1928,Uehling:1933}. 
The BUU equation is semiclassical in character, and similarly to TDDFT, is an equation for the one-body density, even 
though it goes well beyond mean field, but it depends on probabilities, a typical feature of classical descriptions. A quantum approach 
should depend on amplitudes in order to describe interference, quantized vortices, quantum turbulence, and entanglement, 
and the generalized TDSLDA is such a framework~\cite{Bulgac:2022}.  
  
\begin{figure}
\includegraphics[width=1.05\columnwidth]{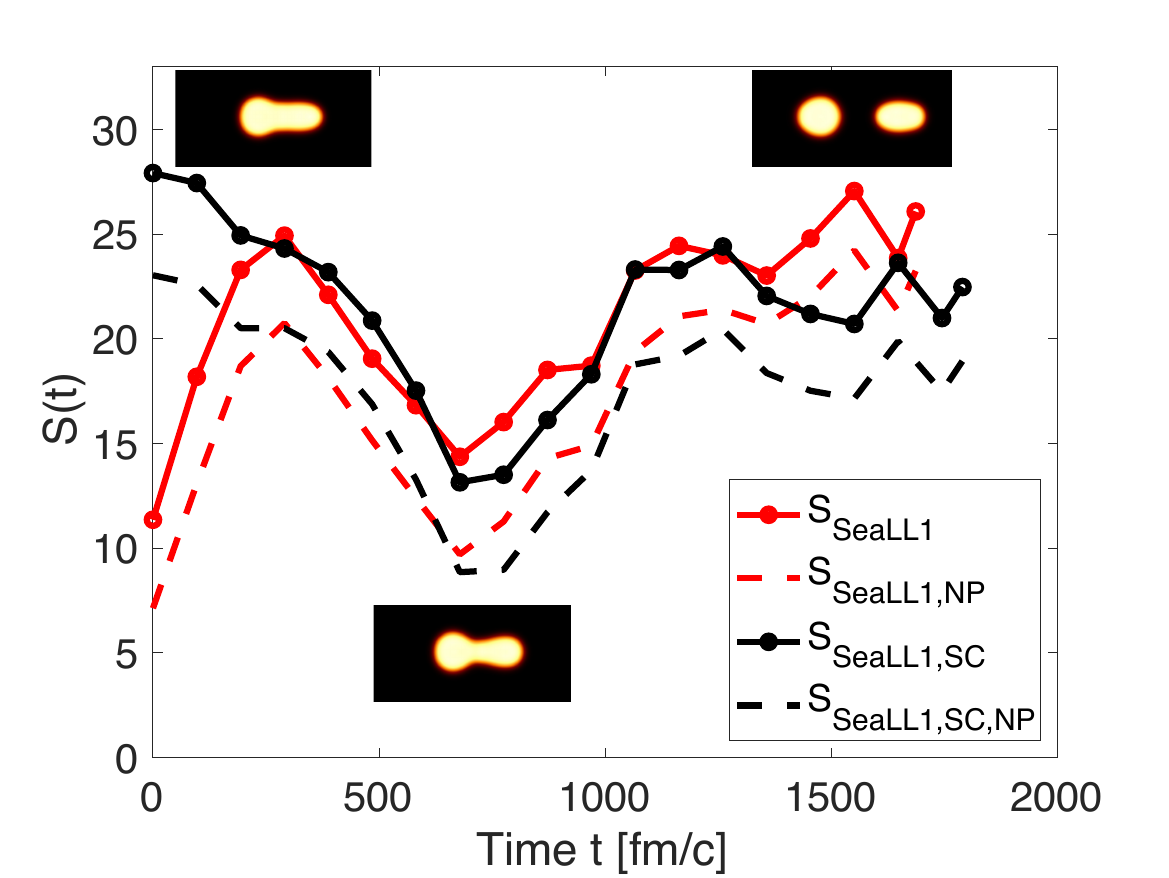}
\caption{ \label{fig:entropy}  
The time-dependence of the entropy $S(t)$ evaluated in the case of the  induced fission of $^{ 235}$U(n,f)
with a low energy neutron as a function of time from the vicinity of the outer saddle point until the two fission fragments are fully separated.
The solid curves correspond are entanglement entropies evaluated without particle projection of the total many-body wave-function, 
while  the dashed curves are obtained after particle projection was performed before the canonical occupation probabilities were evaluated.
(For red curves  see Ref.~\cite{Bulgac:2022c}.) 
The nuclear shapes obtained in TDSLDA during the time evolution are shown at 0, 675, and 1650 fm/c,
from Ref.~\cite{Bulgac:2022c}. }
\end{figure}  
  
A still unresolved problem in theoretical physics is how to characterize the complexity of a many-body wave function. 
The simplest many-body wave function for fermions is a Slater determinant and the introduction of correlations leads to many-body wave 
functions which are sums of many Slater determinants, as in the case of shell-model calculations, where the size of the Hilbert space 
can reach billions~\cite{Johnson:2018}. However, the number of Slater determinants in an expansion of a many-body wave function depends 
exponentially on the size and type of the single-particle basis used and a characterization of the complexity a many-body wave 
function by the dimensionality of the many-body Hilbert space is thus ill-defined. However,
the complexity can be quantified for any quantum state $|\Phi\rangle $ by evaluating the 
orbital entanglement/quantum Boltzmann
entropy~\cite{Nordheim:1928,Uehling:1933,Bulgac:2022,
Horodecki:2009,Haque:2009,Eisert:2010,Boguslawski:2014,Robin:2021}
\begin{eqnarray} 
S = &- g\sum_k n_k\ln n_k 
- g\sum_k  [1-n_k]\ln[1-n_k],\label{eq:ent}
\end{eqnarray}
where $g$ is the spin-isospin degeneracy, $\sum$ implies summation over 
discrete and integral over continuous variables and $n_k$ are the canonical occupation probabilities. 
This orbital entanglement entropy vanishes for a Slater determinant and is positive for any other many-body wave function 
and reaches its maximum value when all $n_k$ are equal to each other. In the limit of a dilute and weakly 
interacting system this entanglement entropy approaches the quantum Boltzmann entropy of that many-fermion system. 
Since the set and the values of the canonical occupation probabilities are independent of the single-particle basis used it 
also allows for a unique definition of the orbital entanglement entropy, which can be used to characterize the complexity 
of the many-body wave function. In the case of a dilute and weakly interacting many-fermion system $S(t)$ as a 
function of time can only increase, $\dot{S}(t)\ge 0$. For a strongly interacting system however the time dependence 
of $S(t)$ is more complicated~\cite{Del-Maestro:2021,Del-Maestro:2022} and the question, what can we learn from it. 
Clearly the orbital entanglement entropy $S(t)$ gives us unique measure on how complicated the many-body wave 
function is at any time during its evolution. $-\ln n_k$ are known in literature as the entanglement spectrum~\cite{Li:2008}, 
play in particular an important role in gauge theories and condensed matter~\cite{Mueller:2022,Li:2008} 
and carry more detailed information about the many-body system than the single number $S(t)$.

We have extracted $S(t)$ in the case of $^{235}$U(n,f) for both particle 
unprojected and particle projected  many-nucleon wave functions within TDSLDA, see Fig.~\ref{fig:entropy}. 
After particle projection the nuclear many-body wave function is a sum over an exponentially 
large number of Slater determinants with fixed $N$ and $Z$ numbers, thus indeed a very complex function. While a fissioning nucleus evolves from 
the top of the outer barrier it develops a neck, which hinders the particle exchange between the two halves of the system. 
Immediately after scission however the two FFs are highly excited,  strongly entangled, and mostly isolated
apart from the long-range Coulomb interaction between them.
Their shapes still evolve in time and the two FFs evolve towards their individual thermal equilibrium and $\dot{S}(t)>0$ as expected, due 
to the presence of $nn$ and $pp$ collisions, which lead to the long momentum tails $n_k=C/k^4$.  For more details see 
Refs.~\cite{Bulgac:2022,Bulgac:2022a,Bulgac:2022c}.
  
\section{Conclusions} 

The only input needed to extract information about the fission dynamics is encoded in only seven parameters, which 
define the nuclear energy density functional for both stationary and time-dependent phenomena and
which are known with good accuracy for decades.
Within  the TDSLDA description of fission we have shown that the nuclear large amplitude collective motion is strongly dissipative, 
extracted TKE, TXE, average FF proton and neutron, have shown how the excitation energy is shared between FFs and how it 
depends on the excitation energy of the compound nucleus, the evolution of the FF shapes after scission, evaluated the FF 
spin distribution  and their correlations,  have insight into the character of the neutrons emitted immediately after scission, 
but before the FFs are fully accelerated, described the prompt neutron multiplicity and their dependence on the energy 
of the compound nucleus, and evaluated the entanglement entropy, which sheds an unparalled light on the complexity of the time-dependent
many-body wave function of a fissioning nucleus and on how its complexity evolves with time. This time-dependent  
microscopic approach has an strong predictive power, without any assumptions and 
uncontrolled approximations and fitting parameters, has already help guide and improve phenomenological models, 
delivers accurate and comprehensive information, and predicts properties which are not accessible in the laboratory, and 
likely will play an important role in helping elucidate the origin of elements, which are formed through fission of nuclei 
which are not possible to study experimentally.  
  
The fission results presented here have been obtained since 2016 in collaboration with I. Stetcu, S. Jin, 
I. Abdurrahman, K. Godbey, K.J. Roche, P. Magierski, and N. Schunck.
The funding  from the US DOE, Office of Science, Grant No. DE-FG02-97ER41014 and
also the support provided in part by NNSA cooperative Agreement
DE-NA0003841 is greatly appreciated. 
This research used resources of the Oak Ridge
Leadership Computing Facility, which is a U.S. DOE Office of Science
User Facility supported under Contract No. DE-AC05-00OR22725.

\providecommand{\selectlanguage}[1]{}
\renewcommand{\selectlanguage}[1]{}

\bibliography{Bulgac}


\end{document}